\documentclass[pdftex,preprint,pra,showpacs,endfloats*]{revtex4} 
\usepackage{natbib}
\usepackage{amssymb}
\usepackage{amsmath}
\usepackage[dvips]{graphicx}
\usepackage{epstopdf}	

\newcommand{\bc}[1]{$\mathbf{#1}$}

\begin{document}

\title{Measurements of the Casimir-Lifshitz force in fluids: the effect of electrostatic forces and Debye screening}
\author{J. N. Munday}
\affiliation{Department of Physics, Harvard University, Cambridge, MA 02138, USA}
\author{Federico Capasso}
\affiliation{School of Engineering and Applied Sciences, Harvard University, Cambridge, MA 02138, USA}
\author{V. Adrian Parsegian and Sergey M. Bezrukov}
\affiliation{National Institutes of Health, Bethesda, MD 20892, USA}
\date{\today}

\begin{abstract}

In this work, we present detailed measurements of the Casimir-Lifshitz force between two gold surfaces (a sphere and a plate) immersed in ethanol and study the effect of residual electrostatic forces, which are dominated by static fields within the apparatus and can be reduced with proper shielding. Electrostatic forces are further reduced by Debye screening through the addition of salt ions to the liquid. Additionally, the salt leads to a reduction of the Casimir-Lifshitz force by screening the zero-frequency contribution to the force; however, the effect is small between gold surfaces at the measured separations and within experimental error. An improved calibration procedure is described and compared to previous methods. Finally, the experimental results are compared to Lifshitz's theory and found to be consistent for the materials used in the experiment.

\end{abstract}

\maketitle

\section{INTRODUCTION}
The prediction of an attractive, quantum electrodynamic force between two electrically neutral metal plates in vacuum has been of great scientific interest since Hendrik Casimir originally proposed the effect in 1948 \cite{Casimir}. A number of reviews describe various aspects of this and other effects resulting from the boundary conditions imposed on the quantized electromagnetic fields \cite{Review0,Review1,Review2}. The early work of Derjaguin \cite{Derjaguin1,Derjaguin2}, Sparnaay \cite{Sparnaay}, and van Blokland and Overbeek \cite{Overbeek} demonstrated the effect; however it was not until the experiment of Lamoreaux \cite{exp1} that a precision measurement of the Casimir force was made. Several additional experiments further confirmed the effect using a variety of techniques \cite{exp2,exp3,exp4}.
	
A more general theory was developed by Lifshitz, Dzyaloshinskii and Pitaevskii, which takes into account the non-ideal reflectivity of real metals, and allows for the calculation of forces between bodies of arbitrary dielectric function separated by a medium, which need not be vacuum \cite{lifshitz1,lifshitz2}. This formalism, which is based on the fluctuation-dissipation theorem, completely describes the van der Waals force at short range \cite{ninham,isra,parse} and simplifies to Casimir's result for ideal metal plates separated by vacuum. 

Recently, there has been interest in the measurement of long-range quantum electrodynamic forces between solids immersed in a fluid to test both the generality of Lifshitz's theory as well as the technological feasibility of incorporating these forces into nano- and micro-devices. It is of theoretical interest to better understand the applicability of the electromagnetic stress tensor to dissipative media used for such calculations \cite{emStress} . From a technological viewpoint, tailoring these forces could lead to methods for reducing stiction in MicroElectroMechanical Systems (MEMS) \cite{MEMS1} and to the development of ultralow friction devices and sensors \cite{Review0}.

Two of the current authors (JNM and FC) recently reported measurements of the Casimir-Lifshitz force between two metal surfaces in a fluid \cite{munday}. In response to criticism that electrostatic effects were not adequately taken into account \cite{GeyerC}, we reported additional experiments that showed that residual electrostatic contributions to the total force were small and that they could not be due to work function differences alone between the metal films but were likely due to stray fields associated with trapped charges \cite{mundayR}.

In this paper, we describe in detail the origin of the electrostatic forces in our experiments, show how they can be reduced, and present detailed experimental procedures for conducting Casimir-Lifshitz force measurements between a metalized sphere and plate separated by fluid.

\section{EXPERIMENTAL SETUP}
\begin{figure}
	\centering
		\includegraphics[width=0.48\textwidth]{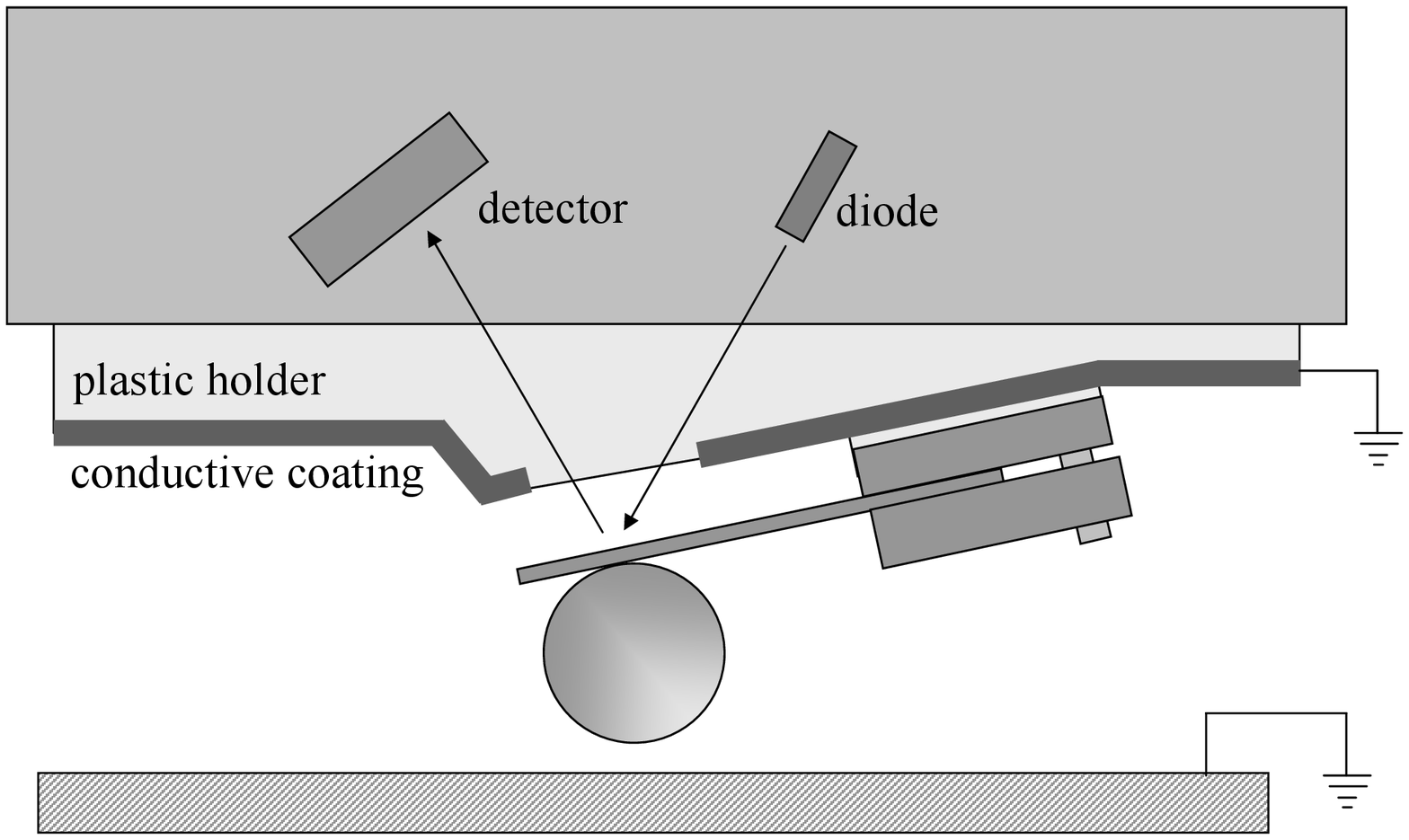}
		\caption{Experimental setup. A polystyrene sphere is attached to an AFM cantilever and coated with gold. A laser beam is directed through a few mm opening in the conductive coating of the cantilever holder and is reflected off the back of the cantilever to monitor its motion.}
	\label{fig:ExpSetup}
\end{figure}

The setup used for this experiment is shown schematically in Fig.~\ref{fig:ExpSetup} and is described in Ref. \cite{munday}. A 39.8 $\mu$m diameter polystyrene sphere is attached to a commercially available cantilever (MikroMasch model CSC38) using silver epoxy. A 5 nm titanium adhesion layer is evaporated onto the sphere and cantilever followed by a 100 nm gold layer. The cantilever chip is then inserted into a commercially available Atomic Force Microscope (AFM) \cite{AFM} with fluid chamber. One major difference in this setup over the one used previously \cite{munday} is the addition of a conductive coating \cite{CPnote} over the plastic cantilever chip holder (Fig.~\ref{fig:ExpSetup}).  This was done to reduce the effect of stray electrostatic fields originating from static charge trapped on the plastic holder. This is discussed in detail in Section IV.

Standard cleaning procedures are performed on all surfaces prior to measurements. The gold plate and fluid cell are ultrasonically cleaned in ethanol for 30 minutes followed by drying in nitrogen airflow. The cantilever chip is rinsed with ethanol without ultrasonic cleaning to avoid damage to the cantilever. For all experiments, the cantilever is completely submerged in ethanol (Sigma-Aldrich), which is filtered through a 0.2 $\mu$m PTFE filter. 
	
Light from a superluminescent diode is reflected off the back of the cantilever and is detected by a four-quadrant photodetector, which is used to monitor the deflection of the cantilever, as in standard AFM measurements. The difference signal between the top two quadrants and the bottom two quadrants is proportional to the vertical deflection of the cantilever. A piezoelectric column within the AFM is used to advance the cantilever and sphere toward the plate, and the piezoelectric column's advance is detected using a linear variable differential transformer \cite{AFM}, which minimizes nonlinearities and hysteresis inherent in piezoelectrics. As the sphere approaches the plate, any force between the two will result in a deflection of the cantilever, which will then be detected in the difference signal from the four-quadrant detector. The sampling rate for data acquisition is chosen so that a minimum of 2 data points per nm are collected and varies from 2 to 8 kHz.
	
The photodetector difference signal is proportional to the distance the cantilever has bent, which obeys Hooke's law: $F_{spring}=-k d_{cantilever}$, where $d_{cantilever}$ is the distance the tip of the cantilever has bent and $k$ is the spring constant of the cantilever. The external force exerted on the sphere, and hence the cantilever, balances the elastic force $F_{spring}$ and is therefore given by: $F_{ext}=\mathcal{C} V_{det}$, where $\mathcal{C}$ is the force constant which is used to convert the photodetector difference signal, $V_{det}$, into a force signal. To determine $\mathcal{C}$, a known force is applied between the plate and the sphere as the sphere approaches the plate, and $\mathcal{C}$ is determined from a fit to this force.

\section{CALIBRATION}
\begin{figure}
	\centering
		\includegraphics[width=0.48\textwidth]{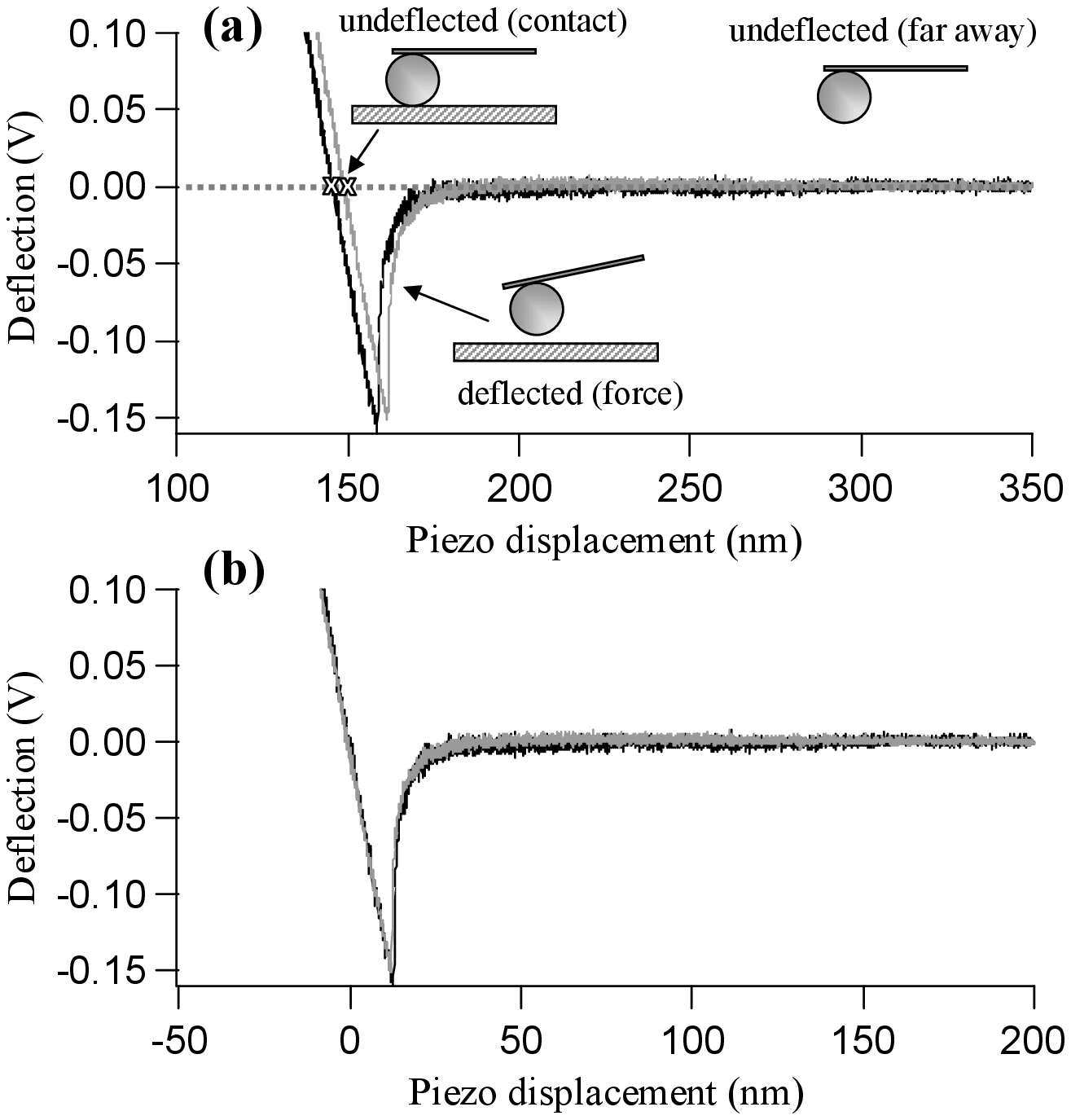}
		\caption{The zero of piezo displacement is defined with respect to contact between the sphere and plate with the cantilever unbent. (a) Two data sets are shown without defining a zero of piezo displacement which is common to both measurements. The x-marker represents the contact between the sphere and plate with the cantilever undeflected and is shown for each of the two runs. The x-markers are offset from each other due to drift between the sample and the sphere as described in the text. (b) By defining the zero of piezo displacement as the point of contact between the sphere and plate when the cantilever is in the undeflected state (see insets (a)), data sets are given a consistent  x-axis and can be averaged.  
		}
	\label{fig:Alignment}
\end{figure}

In order to average the data collected from consecutive runs (typically $\sim$50 data sets are acquired), vertical drift between the sample and the cantilever, which is common in AFM measurements \cite{AFMNote}, is compensated by defining the zero of piezo displacement with reference to the surface of the plate as described below. Figure~\ref{fig:Alignment}(a) shows the photodetector difference signal versus piezo displacement for two runs with a piezo velocity of 45nm/s. The scans were performed approximately 1 min apart and show vertical drift, which is typical of all scans. For large piezo displacements, there is no force between the sphere and the plate. This corresponds to no deflection of the cantilever and $V_{det}=0$. As the sphere approaches the plate, an attractive interaction between the sphere and plate results in a bending of the cantilever until contact is made. At contact, the cantilever is bent. As the piezo column continues to advance, the cantilever unbends (see Fig.~\ref{fig:Alignment}(a)). The point at which the sphere is in contact with the surface and the cantilever is in the unbent state corresponds again to $V_{det}=0$. We define this contact point between the sphere and the plate, when the cantilever is unbent, to be a piezo displacement of zero. The experimental points, which were originally offset horizontally by 3 nm, now coincide at the point of contact (Fig.~\ref{fig:Alignment}(b)). In this way, a large number of data sets from different runs can be consistently averaged.

The photodetector difference signal is proportional to the force exerted on the cantilever plus an additional offset, which is linear with piezo displacement but independent of the actual surface separation. This linear contribution to the signal is due to the relative motion of the cantilever with respect to the laser \cite{munday}. We can write the total force signal as:
\begin{equation}
F_{total}(d,v)=F_{0}(d)+F_{hydro}(d,v)+A d+B,
\label{ForceTot}
\end{equation}
where $F_{0}(d)$ is the Casimir-Lifshitz force (and possibly other non-velocity dependent forces, e.g. electrostatic forces as discussed in the next section), $F_{hydro}(d,v)$ is the hydrodynamic force, and $A$ and $B$ are constants. The hydrodynamic force between a sphere of radius $R$ and a plate separated by a distance $d$, in the limit $R\gg d$, is given by \cite{hydro1,hydro2}: 
\begin{equation}
F_{hydro}(d,v)=-\frac{6\pi \eta v}{d} R^2,
\label{ForceHydro}
\end{equation}
\noindent where $\eta$ is the fluid viscosity and $v$ is the velocity of the sphere relative to the plate (the sign of the velocity is taken to be negative as the sphere approaches the plate). The fact that the sphere is attached to a cantilever does not influence this force because the distance between the cantilever and the plate is large compared to the separation between the sphere and the plate. Thus, the hydrodynamic drag on the cantilever results in only a constant offset in the force signal, which does not depend on the surface separation.

The fluid viscosity, $\eta$, is determined using a Falling Ball Viscometer (Gilmont Instruments) at $21.0\,^{\circ}\mathrm{C}$. The viscometer is calibrated to the data of Ref .\cite{visc} using deionized H$_{2}$O. For ethanol, we find $\eta=1.17 \pm 0.06 \textrm{ mPa s}$ by averaging 10 sets of data. For ethanol solutions with salt (see Section IV), we find $\eta=1.19 \pm 0.08 \textrm{ mPa s}$ and $\eta=1.19 \pm 0.02 \textrm{ mPa s}$ for NaI concentrations of 0.3 mM and 30 mM, respectively. 

\begin{figure}
	\centering
		\includegraphics[width=0.48\textwidth]{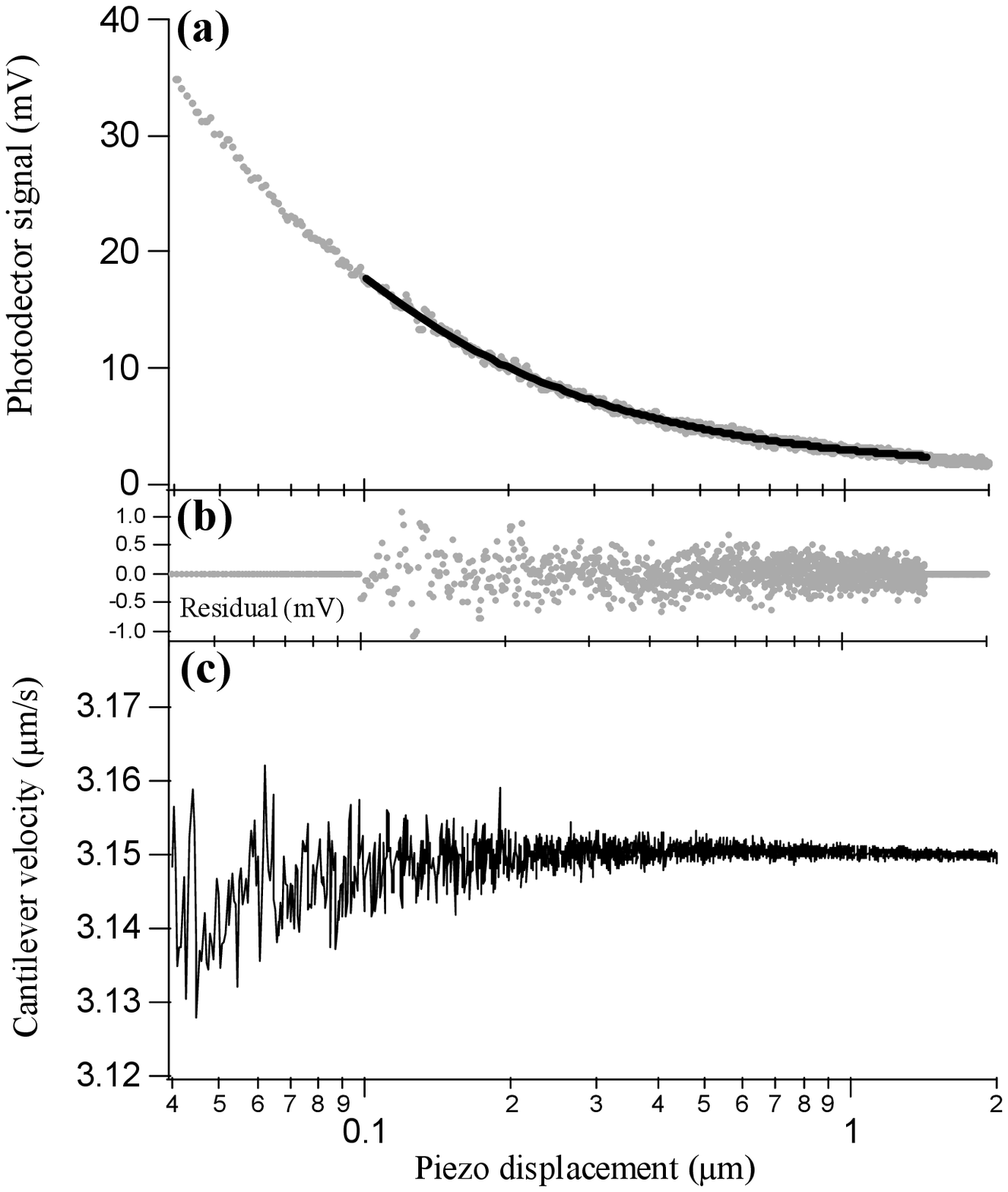}
		\caption{The hydrodynamic force is used for cantilever calibration by a fit to the raw deflection vs. piezo displacement data. (a) Dots are the deflection data corresponding to an effective piezo velocity of $3.150\mu m/s$ using the calibration methods described in Section III. Solid line is a least-squares fit to the data. (b) The residual shows no systematic error to the fit. (c) The cantilever velocity is constant over the distance range used for the calibration (0.1-1.5$\mu$m).}
	\label{fig:Calibration}
\end{figure}

By performing measurements at different piezo velocities and then combining the results, we can determine the various contributions to the total force independently. To determine the hydrodynamic force at a velocity $v$, we make two measurements: one at $v_{1}$ and one at $v_{2}$, such that $v= v_{2}-v_{1}$. Subtracting the results, we get, from Eq. \ref{ForceTot}:
\begin{equation}
F_{total}(d,v_{2})-F_{total}(d,v_{1})= F_{hydro}(d,v).
\end{equation}
\noindent Similarly, $F_{0}(d)+A d+B$ is determined by performing a third measurement at a velocity 2$v_{1}$ and combining the data as:
\begin{equation}
2 F_{total}(d,v_{1})-F_{total}(d,2v_{1})= F_{0}(d)+ A\,d+B.
\label{ForceNH}
\end{equation}
\noindent The linear contribution $(A\,d+B)$ to the total force signal is measured with the sphere far away from the surface so that $F_{0}(d)$ is determined.
	
To convert the photodetector difference signal into a force signal, calibration is performed using the hydrodynamic force \cite{munday,HydroCal}. Figure~\ref{fig:Calibration}(a) shows the photodetector signal versus piezo displacement for the hydrodynamic force with a velocity of 3.150 $\mu$m/s. Grey dots in Fig.~\ref{fig:Calibration}(a) correspond to the average of 51 runs using the methods described above at two different velocities, $v_{2}=3.600\ \mu m/s$ and $v_{1}=0.450\ \mu m/s$. The bending of the cantilever $d_{cantilever}$ is taken into account by adding $d_{cantilever}$, as determined from the deflection signal, to the piezo displacement prior to combining the data obtained with different piezo speeds. The force constant $\mathcal{C}$ and the actual sphere-plate separation at contact $d_{0}$ are determined by fitting the cantilever deflection data to Eq. \ref{ForceHydro} with $d=d_{piezo}+d_{cantilever}+d_{0}$, where $d_{piezo}$ is the piezo displacement and $d_{cantilever}$ is the amount the cantilever has bent. For two ideally smooth surfaces, $d_{0}$=0; however, once contact is made for surfaces with nano-metric roughness, the peaks in surface roughness will preclude further advancement of the surfaces toward each other. $d_0$ is a constant that takes this into account. The fit (solid line in Fig.~\ref{fig:Calibration}(a)) is performed for piezo displacements between 0.1 and 1.5 $\mu$m, which leads to $\mathcal{C}=14.5\pm 0.1\textrm{ nN/V}$ and $d_{0}=12\pm 1\textrm{ nm}$. The residual shows no systematic error in the least-squares fit (Fig. \ref{fig:Calibration}(b)). Figure~\ref{fig:Calibration}(c) shows that the cantilever velocity is nearly constant as a function of piezo displacement. At the smallest separations the velocity becomes slightly reduced due to the bending of the cantilever caused by the force. 

\begin{figure}
	\centering
		\includegraphics[width=0.48\textwidth]{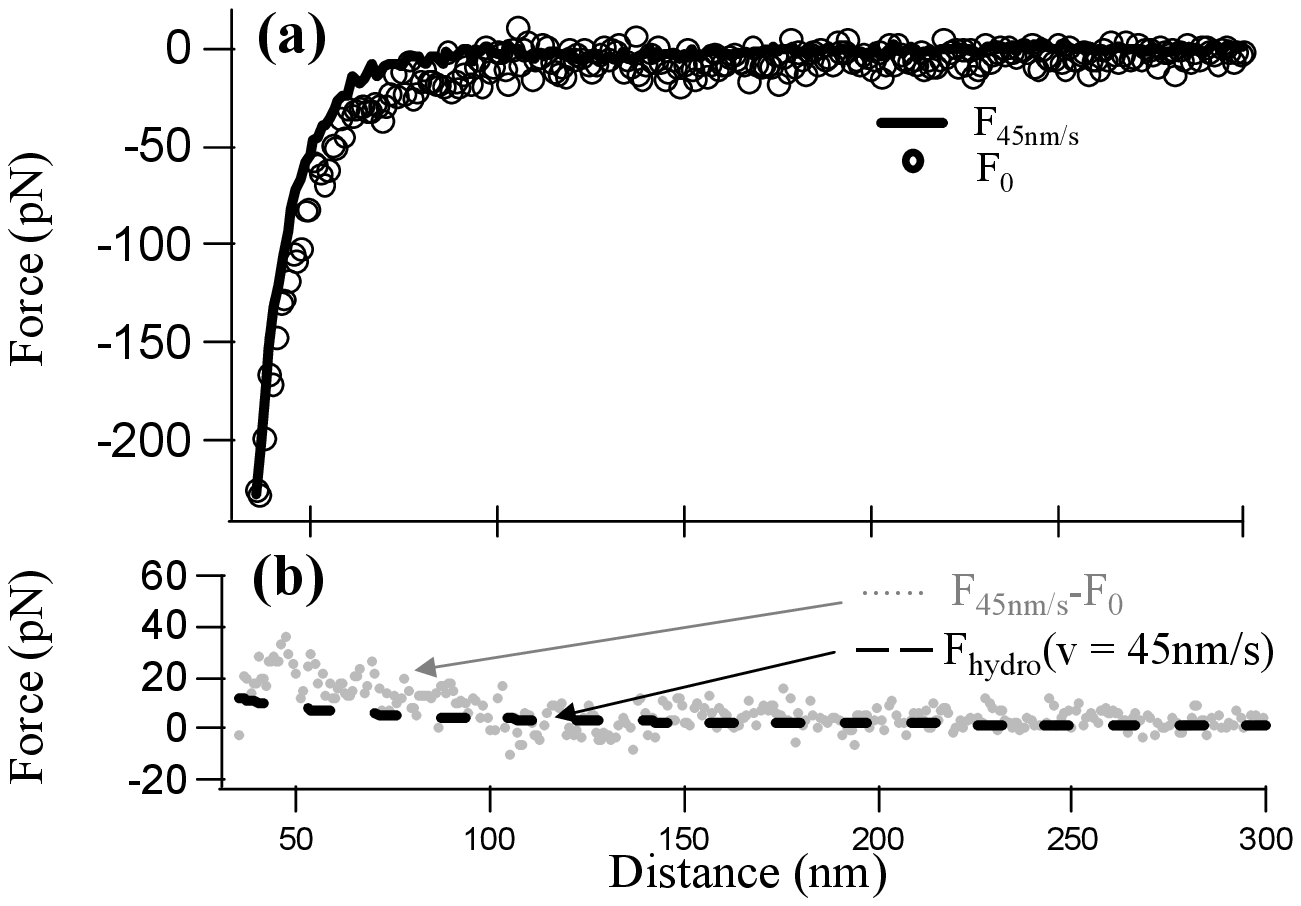}
		\caption{Two methods are compared to determine the Casimir-Lifshitz force and are found to be in reasonable agreement. (a) Solid line corresponds to the force determined using a slow piezo velocity (45 nm/s), where the hydrodynamic force is small. Circles correspond to force data obtained using the method described in Section III. (b) Force difference is calculated between these two methods and is comparable to the residual hydrodynamic force resulting from the piezo velocity of 45 nm/s.}
	\label{fig:MethodCompare}
\end{figure}

Varying the fit range for the data has little effect on the calibration. The fit range is chosen so that a majority of the data is above the noise level and that the modification to the cantilever velocity due to bending is small. Fits are also performed over a slightly larger range and give rise to a modification of $\mathcal{C}$ by $\sim$2\% and $d_{0}$ by $\sim$3.5 nm (Table \ref{tab1}) when the fit range incorporates data where the velocity is not constant. Table \ref{tab1} also shows that a fit at slightly larger separations results in negligible change to $\mathcal{C}$ and $d_{0}$. We note that in addition to the uncertainty in the force due to the calibration constant $\mathcal{C}$, there is an additional uncertainty of of 2-7\% in the force due to the uncertainty in $\eta$.

\begin{table}
\begin{tabular}{|c|c|c|}
\hline
Fit range ($\mu$m) & $\mathcal{C}$ (nN/V) & $d_{0}$ (nm) \\ \hline
$0.05-1.00$ &$14.2\pm0.1$ & $15.5\pm0.5$ \\ \hline
$0.10-1.50$ & $14.5\pm0.1$ & $12\pm1$ \\ \hline
$0.15-1.50$ & $14.5\pm0.1$ & $12\pm2$\\ \hline
\end{tabular}
\caption{Values of the fit parameters obtained using different fit ranges for the deflection data of the hydrodynamic force.}
\label{tab1}
\end{table}
	
The Casimir-Lifshitz force between the two gold surfaces in ethanol is determined using the methods described above (Eq.\ref{ForceNH}) with $v_1=450\textrm{ nm/s}$. The circles in Fig.~\ref{fig:MethodCompare}(a) show the experimental results averaged for 51 runs. The solid line in Fig.~\ref{fig:MethodCompare}(a) is the value of the force obtained with the same sphere using the method of Ref. \cite{munday}, which corresponds to reducing the piezo velocity to a small value (45 nm/s) rather than subtracting the data for different speeds. Figure~\ref{fig:MethodCompare}(b) shows the difference in force using these two methods. The results using a slow piezo velocity appear to be $\sim$20 pN weaker at a distance of 50 nm. This can partially be understood by the residual hydrodynamic force (dashed line), which is still present in this data; however, 20 pN is also comparable to the spread in the data. Finally, the advantage of the method described in this paper is that the drift in the sphere-plate separation is reduced by an order of magnitude because the measurement time is shorter at larger piezo velocities.

\section{ELECTROSTATIC FORCES}
The effect of residual electrostatic forces has been a concern since the earliest experiments to measure the Casimir effect \cite{Sparnaay,Overbeek}. Ideally, there would be no electrostatic force between two identical metals connected through a common ground; however, in real world experiments, two metallic films will have variations in their work functions, which lead to residual electrostatic forces. Variations in the work function over the surface of a metal can lead to position dependent electrostatic forces and patch potentials \cite{Parker1962-948}, which cannot easily be compensated. Further, stray charges and fields from the surroundings can lead to additional electrostatic forces. To quantify such effects, a series of experiments are conducted to determine the contribution of electrostatic forces to the total measured force in our experiment.
	
\begin{figure}
	\centering
		\includegraphics[width=0.48\textwidth]{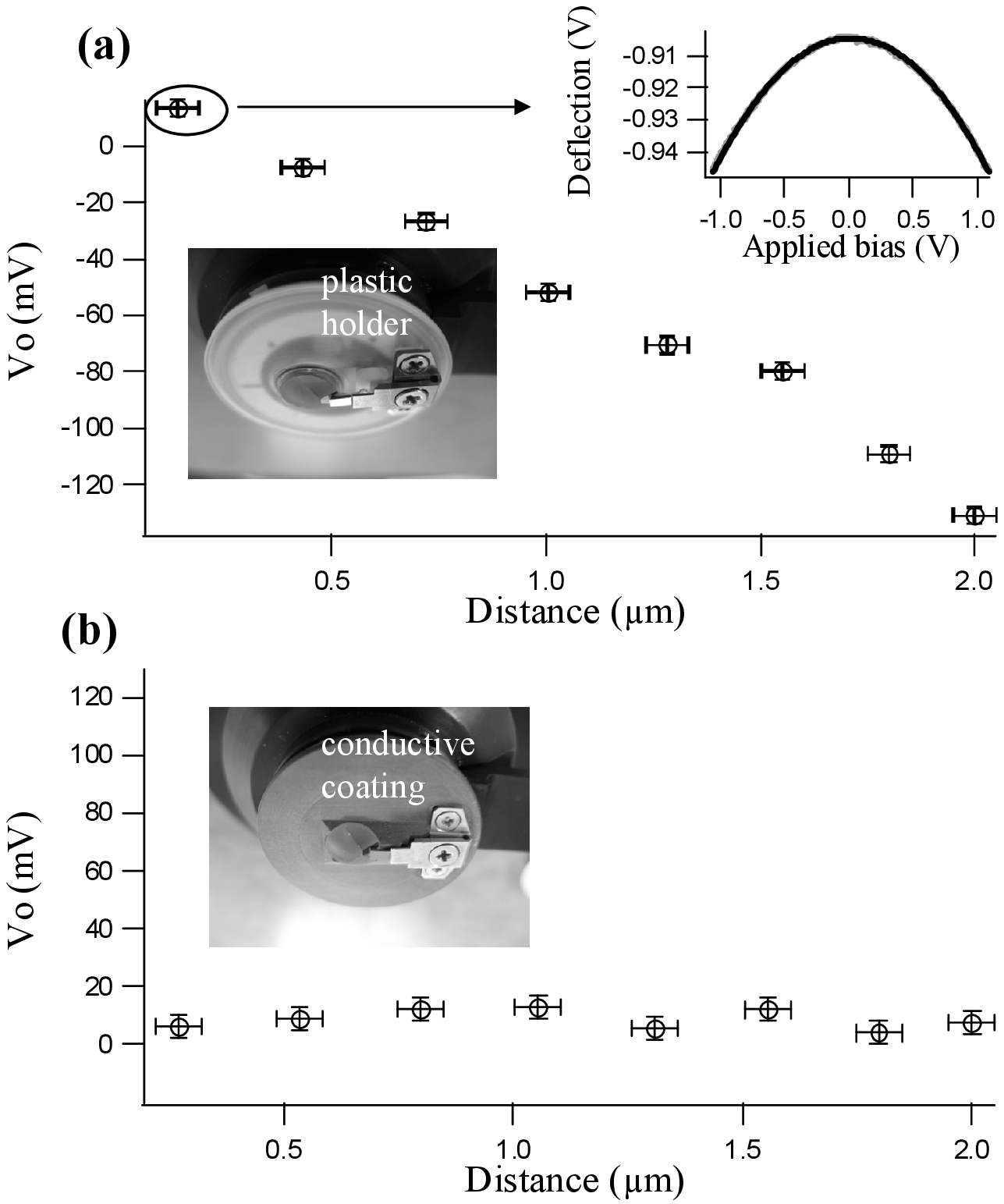}
		\caption{Variation in $V_{0}$, the applied voltage needed to minimize the electrostatic force, with distance is found to result from static charge on the cantilever chip holder. (a) The applied tip voltage that yields minimum deflection of the cantilever, $V_{0}$, is plotted for various surface separations. With the plastic holder, $V_{0}$ is found to vary with surface separation. (Inset) Typical deflection versus the applied voltage on the tip relative to the grounded sample. (b) Same as (a) but with a conductive coating on the cantilever chip holder. The variation in $V_{0}$ with distance is minimized.}
	\label{fig:V0}
\end{figure}

For measurements of the Casimir force in vacuum or air, a bias voltage is applied to the sphere, while the plate is grounded, to determine the voltage, $V_{0}$, needed to compensate and minimize the electrostatic force \cite{Overbeek}. Using a plastic cantilever chip holder without a conductive coating, $V_{0}$ appears to vary with surface separation by $\sim$140 mV over $\sim$2 $\mu$m (Fig.~\ref{fig:V0}(a)). We attribute this variation to stray charge on the plastic cantilever chip holder. A conductive coating \cite{CPnote} is applied to the holder and grounded (Fig.~\ref{fig:ExpSetup}). This reduces the value of $V_{0}$ and its distance dependence (Fig.~\ref{fig:V0}(b)). These results support our original claim that stray charges and fields were the dominant source of residual electrostatic forces in our previous experiment \cite{munday,mundayR}. For the experiments presented in this paper, the holder with the conductive coating is used and the sphere and plate are grounded unless otherwise stated.

\begin{figure}
	\centering
		\includegraphics[width=0.48\textwidth]{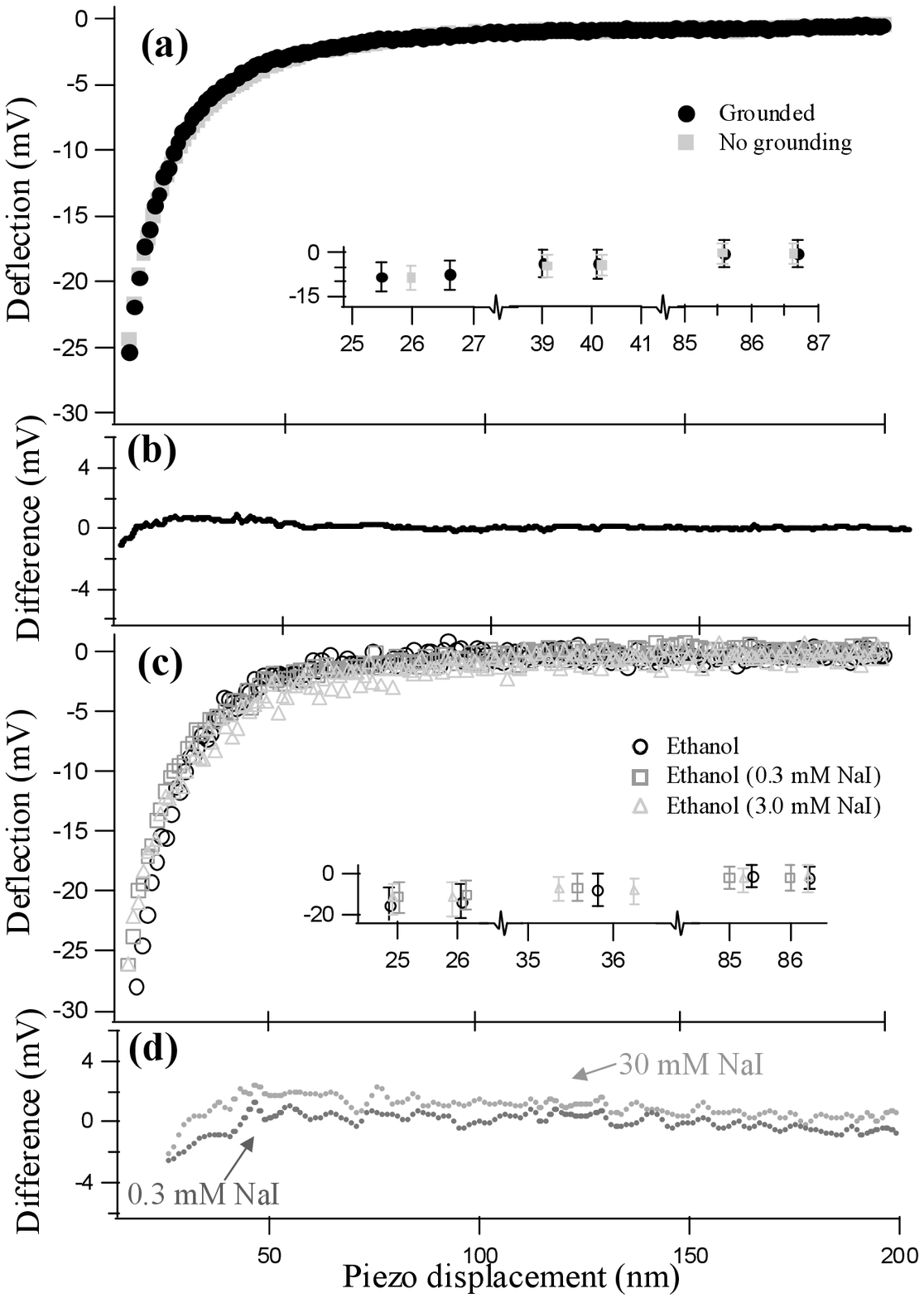}
		\caption{The effect of the grounding and salt concentration on the total force is small. (a) Cantilever deflection with and without grounding the sphere and plate in ethanol. (b) Difference in deflection signal between data with and without grounding. (c) Deflection data with various salt (NaI) concentrations in ethanol. (d) Deflection difference between the data with various salt concentrations and no added salt. Insets: Selected points from the main graph show vertical error bars that represent standard deviation for 51 measurements. No horizontal error bars are shown; however, the uncertainty in $d_0$ from the hydrodynamic fit is 1 nm.}
	\label{fig:ESTests}
\end{figure}

The effect of grounding the sphere and plate is investigated during force measurements in fluids and found not to affect the net force for piezo displacements above 30nm (Fig.~\ref{fig:ESTests}(a,b)). It is likely that the $V_{0}$ measured using the conductive cantilever holder results from the work function difference between the sphere and the plate \cite{WFNote}. The electrostatic force in ethanol due to the work function difference can be written as:
\begin{equation}
F_{ES}=-\frac{\pi R \epsilon_{ethanol} \epsilon_{0} V_{0}^2}{d} e^{-d/\lambda_D},
\label{ES}
\end{equation}
\noindent where $\epsilon_{ethanol}=24.3$ is the dielectric constant of ethanol and $\lambda_D$ is the Debye screening length. For $\lambda_D$ between 20 nm and 100 nm (which is typical for commercially available ethanol \cite{DebyeNote}), the electrostatic force at 30 nm is between 6 pN and 21 pN.

To further ensure that electrostatic forces are negligible, varying amounts of sodium iodide (NaI) were added to the ethanol to decrease the Debye screening length, $\lambda_D$. Figure 6(c) shows the cantilever deflection versus piezo displacement for three difference concentrations of NaI in ethanol: (1) no added NaI, (2) 0.3 mM NaI corresponding to $\lambda_D=10\textrm{ nm}$, and (3) 30 mM NaI corresponding to $\lambda_D=1\textrm{ nm}$. The difference in the deflection signal between the data acquired with no added salt and the data acquired with salt is within the experimental error for the measured piezo displacements (Fig.~\ref{fig:ESTests}(d)); however, small variations are expected in the Casimir-Lifshitz force as a result of the added salt as discussed in Sections V and VI.
	
\begin{figure}
	\centering
		\includegraphics[width=0.48\textwidth]{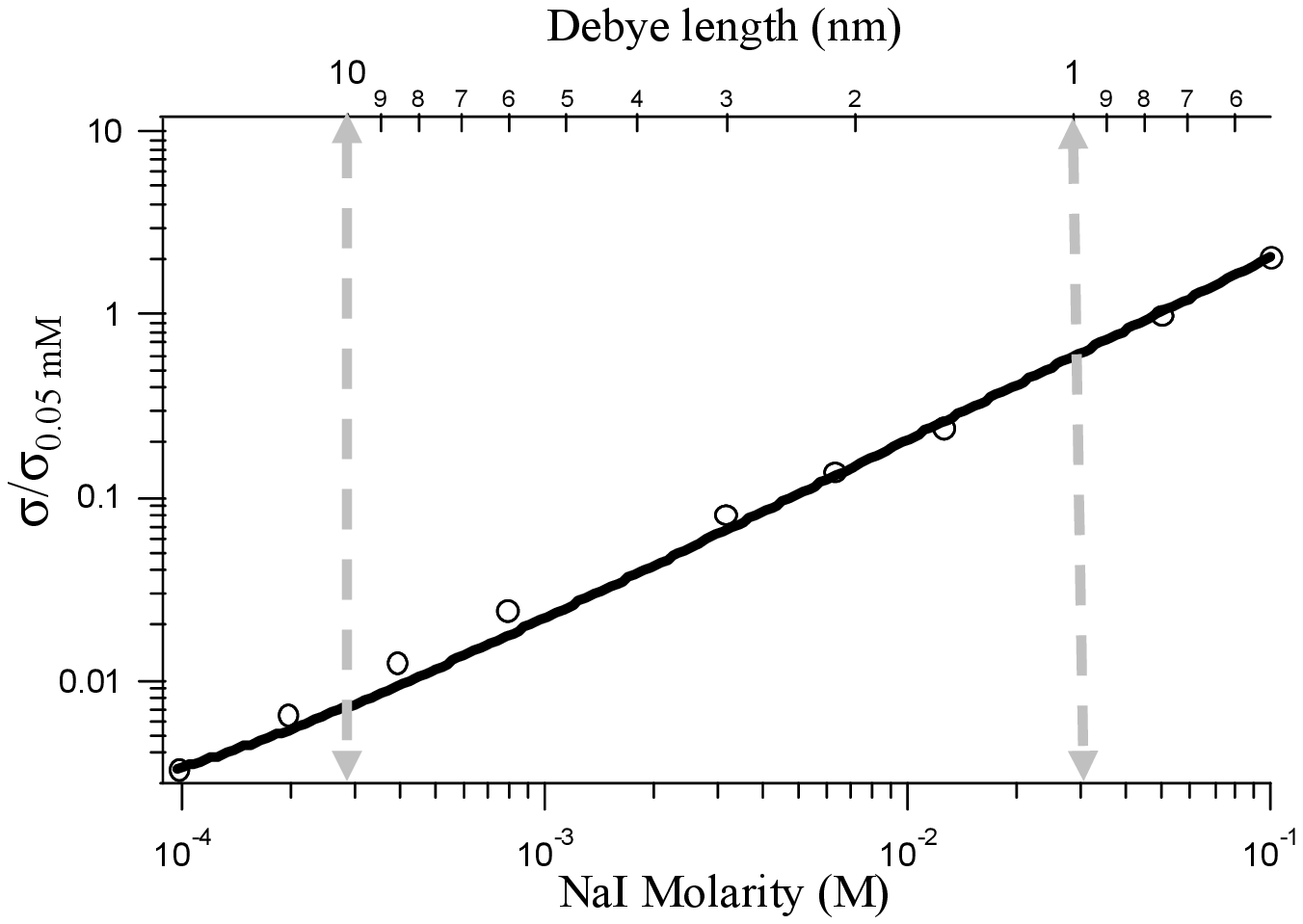}
		\caption{Conductivity is found to vary nearly linearly with molarity. Circles represent measurements, and the solid line is a linear fit to the data (log-log scale). Grey arrows represent the two largest salt concentrations used for force measurements.}
	\label{fig:Conductivity}
\end{figure}

The specific conductivity of the solution is found to increase nearly linearly with increasing salt concentration. Figure~\ref{fig:Conductivity} shows conductivity data for NaI in ethanol (circles), normalized to the value at a concentration of 0.05 M, versus molar concentration obtained using a bench conductivity meter (VWR). The solid line is a linear fit to the data showing that the fluid/salt mixture has not saturated, consistent with Kohlrausch's law \cite{Kohl1, Kohl2}. The grey arrows indicate salt concentrations used for experiments and corresponding screening lengths.
\section{THEORY}
The experimental data are compared to Lifshitz's theory for a gold sphere of radius $R$ and a gold plate separated by a distance $d$ in ethanol. For this configuration at temperature $T$, Lifshitz's theory can be written as (see for example Ref. \cite{Review1}):

\begin{equation}
\begin{split}
F_{C\textrm{-}L}(d)=&k_{B} T R \sideset{}{'}\sum_{m=0}^\infty\int_{k=0}^\infty k [ \ln ( 1- r_{3,1}^{TE} r_{3,2}^{TE} e^{-2k_{3}d}) 
\\& + \ln ( 1- r_{3,1}^{TM} r_{3,2}^{TM} e^{-2k_{3}d}) ] dk
\end{split}
\label{FCL1}
\end{equation}

\noindent where $k_{B}$ and $c$ are the usual fundamental constants, and the primed summation gives half weight to the $m=0$ term. $r_{3,i}^{p}$ is the reflection amplitude at the interface between material $i$ and material 3 for radial wave vectors of magnitude $k$ and polarization $p$, given by

\begin{equation}
r_{3,i}^{TE}=\frac{k_{i}-k_{3}}{k_{i}+k_{3}}
\qquad \text{and} \qquad 
r_{3,i}^{TM}=\frac{k_{i}\epsilon_{3}-k_{3}\epsilon_{i}}{k_{i}\epsilon_{3}+k_{3}\epsilon_{i}}
\end{equation}

\noindent where

\begin{equation}
k_{i}=\sqrt{k^2+\frac{\epsilon_{i}\xi^2}{c^2}}.
\end{equation}

\noindent $\epsilon_{1}$, $\epsilon_{2}$, and $\epsilon_{3}$ are the dielectric functions of the sphere, plate, and intervening medium, respectively, evaluated at imaginary frequencies $i\xi =i\frac{2\pi k_{B} T}{\hbar}m$ according to:

\begin{equation}
\epsilon_{i}(i\xi)=1+\frac{2}{\pi}\int_{x=0}^\infty \frac{x\ \textrm{Im}[\epsilon_{i}(x)]}{x^2+\xi^2}dx.
\end{equation}

The dielectric function for gold is obtained using the optical data of Palik \cite{Palik} for $\omega$ from 0.125 to 9184 eV and the Drude model, 

\begin{equation}
\epsilon(\omega)=1-\frac{\omega_{p}^2}{\omega (\omega+i\gamma)},
\label{Drude}
\end{equation}

\noindent where $\omega_{p}=7.50$ eV and $\gamma=0.061$ eV for $\omega <0.125$ eV. The Drude parameters are obtained from Ref. \cite{sample1}, where the data of Palik were extrapolated to lower frequencies by a fit to Eq. \ref{Drude} for the mid-infrared. For ethanol, a two-oscillator model is used \cite{2osc1,2osc2}:

\begin{equation}
 \epsilon(i\xi)=1+{C_{IR}\over 1+\left({\xi\over \omega_{IR}}\right)^2}+{C_{UV}\over 1+\left({\xi\over \omega_{UV}}\right)^2}
\label{twoosc}
\end{equation}

\noindent where $\omega_{IR}=6.60\times 10^{14}$ rad/s and $\omega_{UV}=1.14\times 10^{16}$ rad/s are the characteristic absorption angular frequencies in the infrared and ultraviolet range, respectively, and $C_{IR}=23.84$ and $C_{UV}=0.852$ are the corresponding absorption strengths. 

It has been shown that the natural variability in the optical properties of materials can give rise to a significant variation in the Casimir force, which can be $>$10\% \cite{sample1,sample2,mundayR}. For this reason, Lifshitz's theory will not completely describe the measured force unless the optical properties of the actual materials used in the experiment are determined over a large frequency range for the sphere, plate and fluid. Such measurements are a significant experimental challenge and have led to the adaption of tabulated data for calculations \cite{sample3}.
	
The salt concentration further modifies Eq. \ref{FCL1} in two ways. First, the salt will partially screen the zero-frequency component of the Casimir-Lifshitz force. To account for this, the zero-frequency contribution in Eq. \ref{FCL1} is replaced by \cite{Review1}:

\begin{equation}
\begin{split}
F_{C\textrm{-}L}|_{m=0}=&\frac{k_{B} T R}{2}\int_{k=0}^\infty k \ln[1
-\Big(\frac{\epsilon_{1} k-\epsilon_{3}\sqrt{k^2+\kappa^2}}{\epsilon_{1} k+\epsilon_{3}\sqrt{k^2+\kappa^2}}\Big)
\\&
\times\Big(\frac{\epsilon_{2} k-\epsilon_{3}\sqrt{k^2+\kappa^2}}{\epsilon_{2} k+\epsilon_{3}\sqrt{k^2+\kappa^2}}\Big) e^{-2 d \sqrt{k^2+\kappa^2}}] dk,
\end{split}
\end{equation}

\noindent where $\kappa=1/\lambda_{D}$. Second, the salt can modify the optical properties of ethanol. As a first order correction to the dielectric function at non-zero frequency, a plasma term \cite{ashcroft} was added to Eq. \ref{twoosc} for ionic species of sodium and iodine. The effect of the salt on the optical properties using this approximation is negligible. 
	
\begin{figure}
	\centering
		\includegraphics[width=0.48\textwidth]{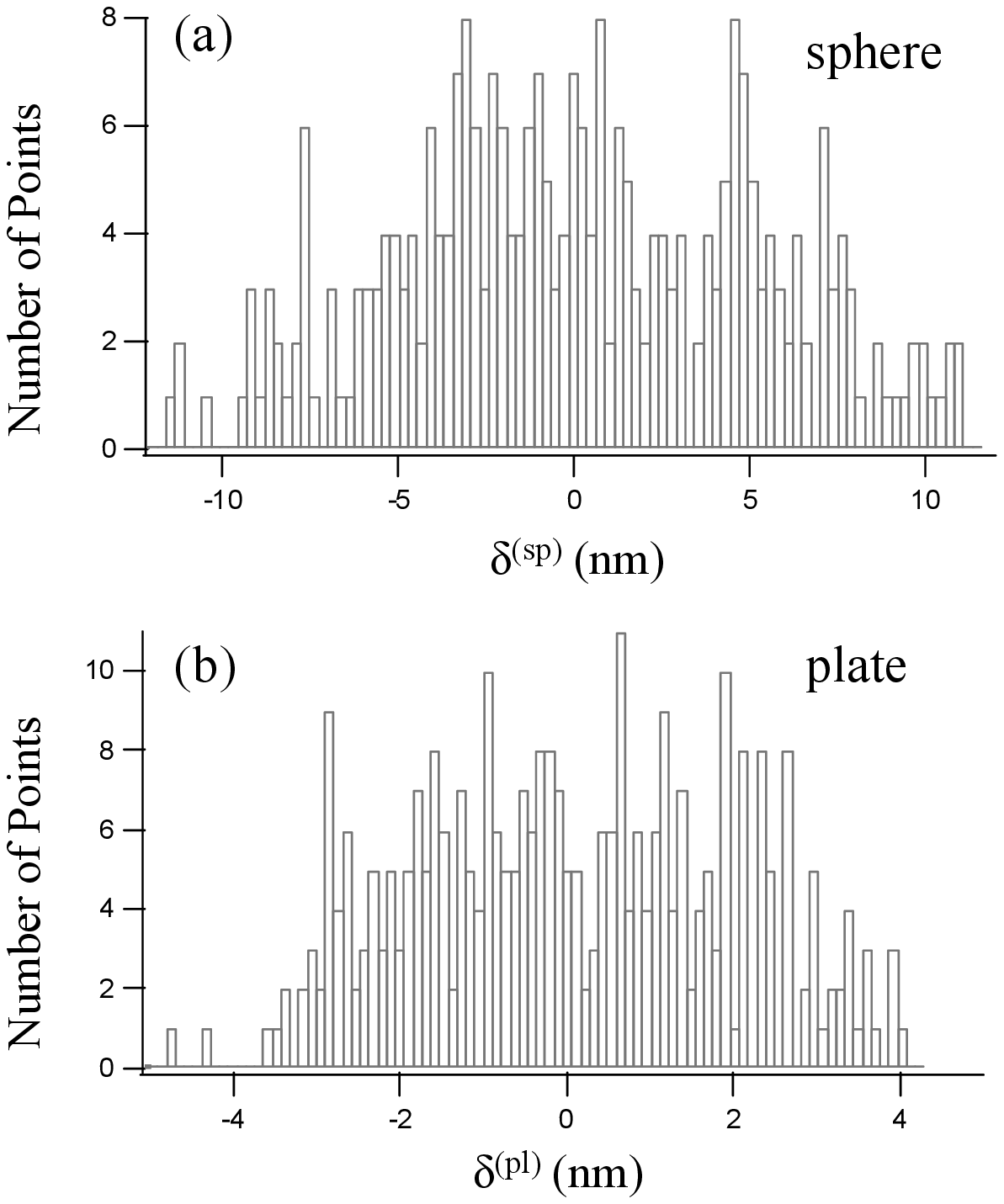}
		\caption{Measurement of the surface roughness of the gold coated sphere (a) and 
gold coated plate (b) used in the experiment. Bar heights represent the number of pixels 
of an optical profiler image in which the height of the sphere (plate) is vertically 
displaced by an amount $\delta^{(sp)}$ ($\delta^{(pl)}$) with respect to an ideally smooth surface. }
	\label{fig:Roughness}
\end{figure}

Surface roughness further modifies the calculation of the force. The total Casimir-Lifshitz force including this correction can be calculated as \cite{rough}:

\begin{equation}
F(d)=\sum_{i,j} \sigma_{i}^{(sp)} \sigma_{j}^{(pl)} F_{C\textrm{-}L}[d-(\delta_{i}^{(sp)}+\delta_{j}^{(pl)})],
\end{equation}

\noindent where  $\sigma_{i}$ is the fraction of the surface area of the sphere (sp) or plate (pl) displaced a distance $\delta_{i}$ from an ideally smooth surface and are measured over a 2 $\mu$m by 2 $\mu$m area using an optical profiler (Fig.~\ref{fig:Roughness}).

\begin{figure}
	\centering
		\includegraphics[width=0.48\textwidth]{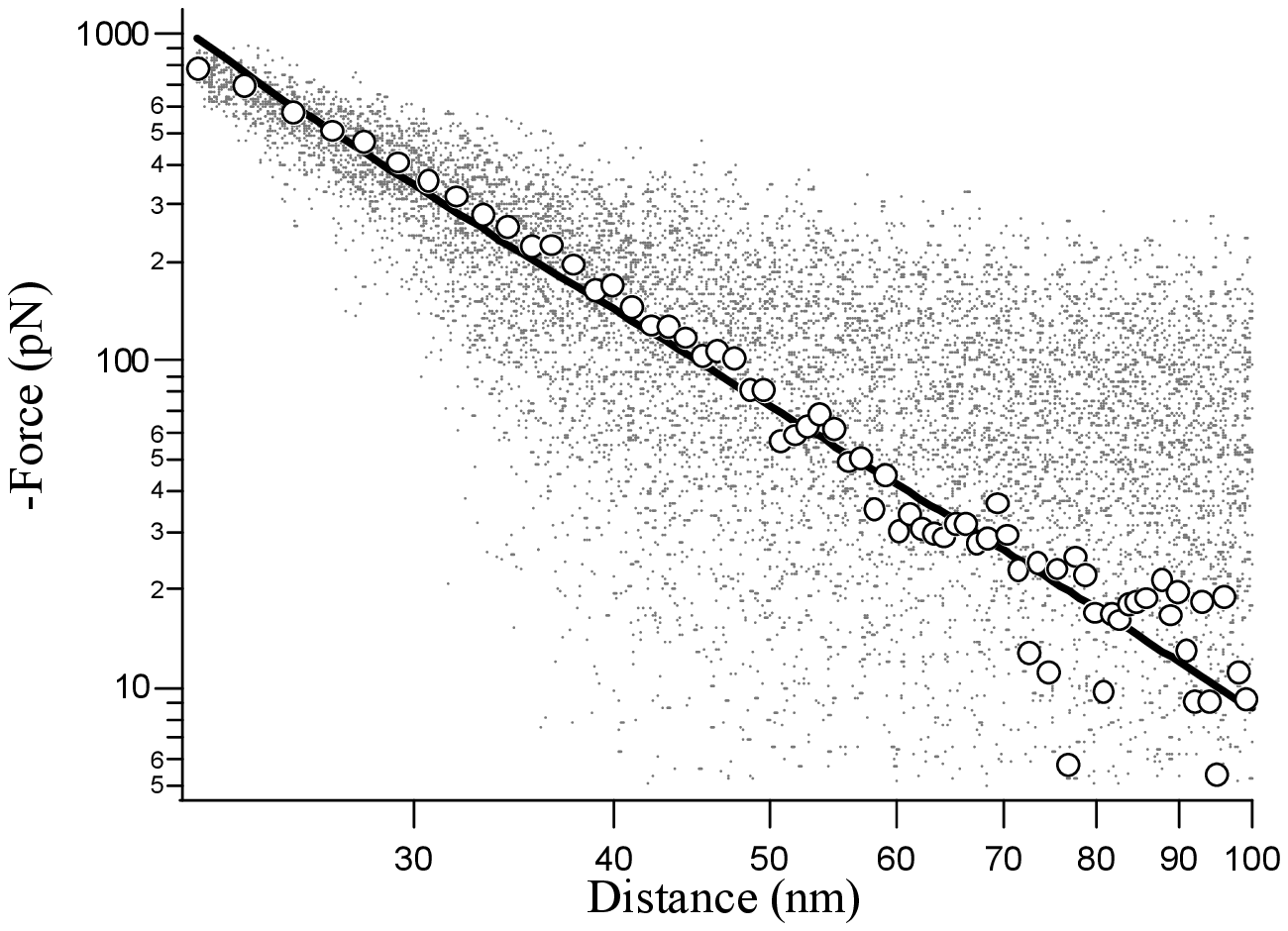}
		\caption{Measured force between a gold sphere and a gold plate 
immersed in liquid ethanol is well described by Lifshitz's theory. Dots represent 
measurements from 51 runs over a distance range of approximately 20-100nm. The increased density of points for larger force values is due to log compression. Also note that not all data points are shown in the figure. In particular, there are many data points that exist below the x-axis in the distance range of 50-100nm that are not displayed because of the log plot. Circles are average values from the 51 data sets. Solid line is Lifshitz's theory.}
	\label{fig:LogForce}
\end{figure}

\section{RESULTS}
\begin{figure}
	\centering
		\includegraphics[width=0.48\textwidth]{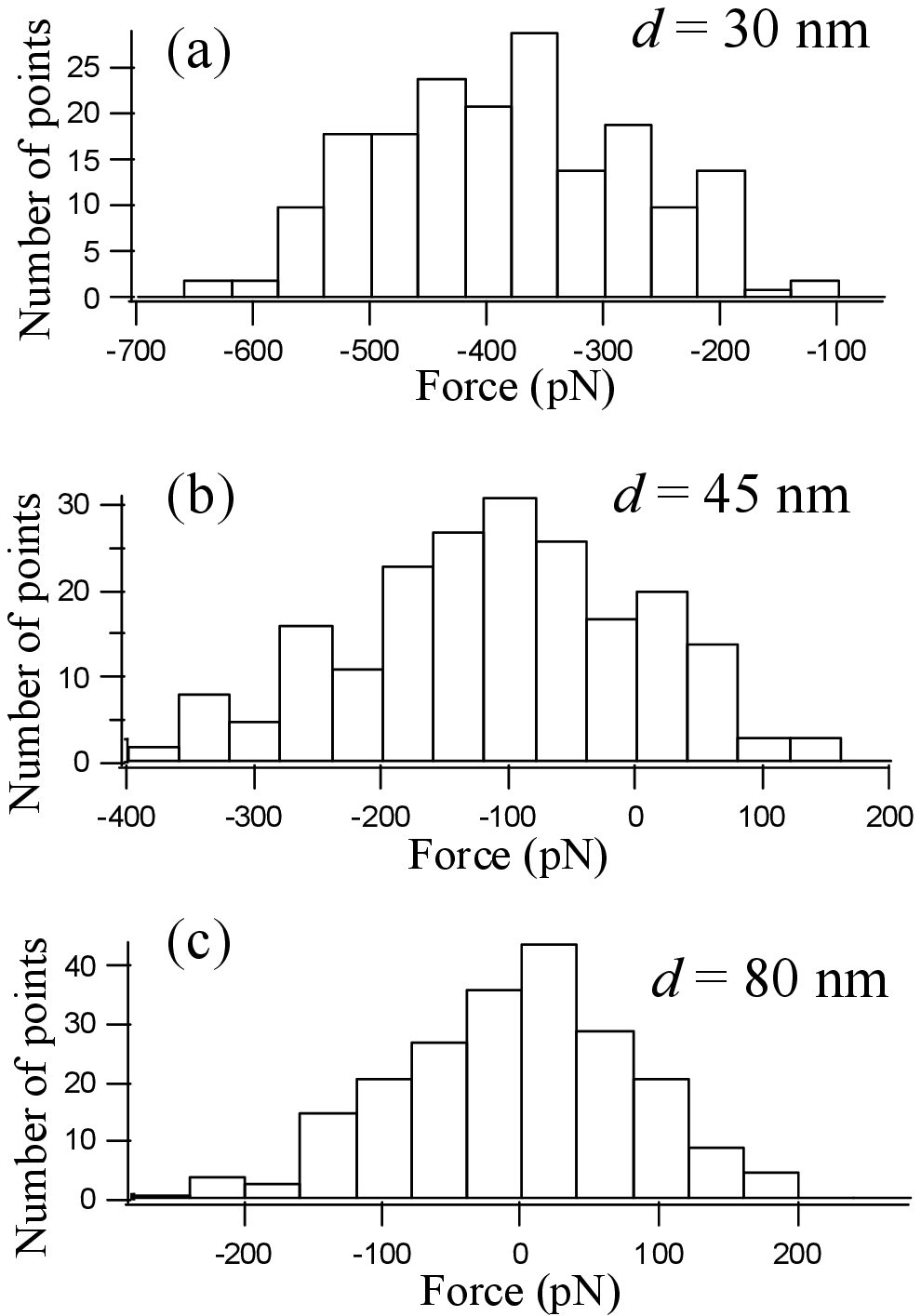}
		\caption{Histograms of the force data show an approximately Guassian 
distribution for data at various surface separations between the sphere and plate: (a) 30 
nm, (b) 45nm, and (c) 80 nm. Force data is collected and combined from 51 runs as 
discussed in the text. Distances are rounded to be in steps of 1 nm.}
	\label{fig:Histogram}
\end{figure}

Figure~\ref{fig:LogForce} shows the Casimir-Lifshitz force in ethanol between the gold-coated sphere and gold-coated plate. The data for 51 runs are shown (dots) along with the average of these data (circles) and the theory described in Section V for ethanol with no added salt (solid line). The theory describes the data well, despite the uncertainties in the optical properties. Note that both the theory and the experimental data follow a $\sim$ $d^{-3}$ dependence for the force in the presented distance range, as expected due to retardation effects (e.g. tables P.1.b.3 and L2.2.B in Ref. \cite{Review1}). Histograms of the force data from the 51 runs show an approximately Gaussian distribution at all separations and no evidence of systematic errors (Fig.~\ref{fig:Histogram}). Deviations between the theory and experiment below 30nm are likely due to the inability of the theory to accurately describe the surface roughness on these scales and the uncertainty in the optical properties.

\begin{figure}
	\centering
		\includegraphics[width=0.48\textwidth]{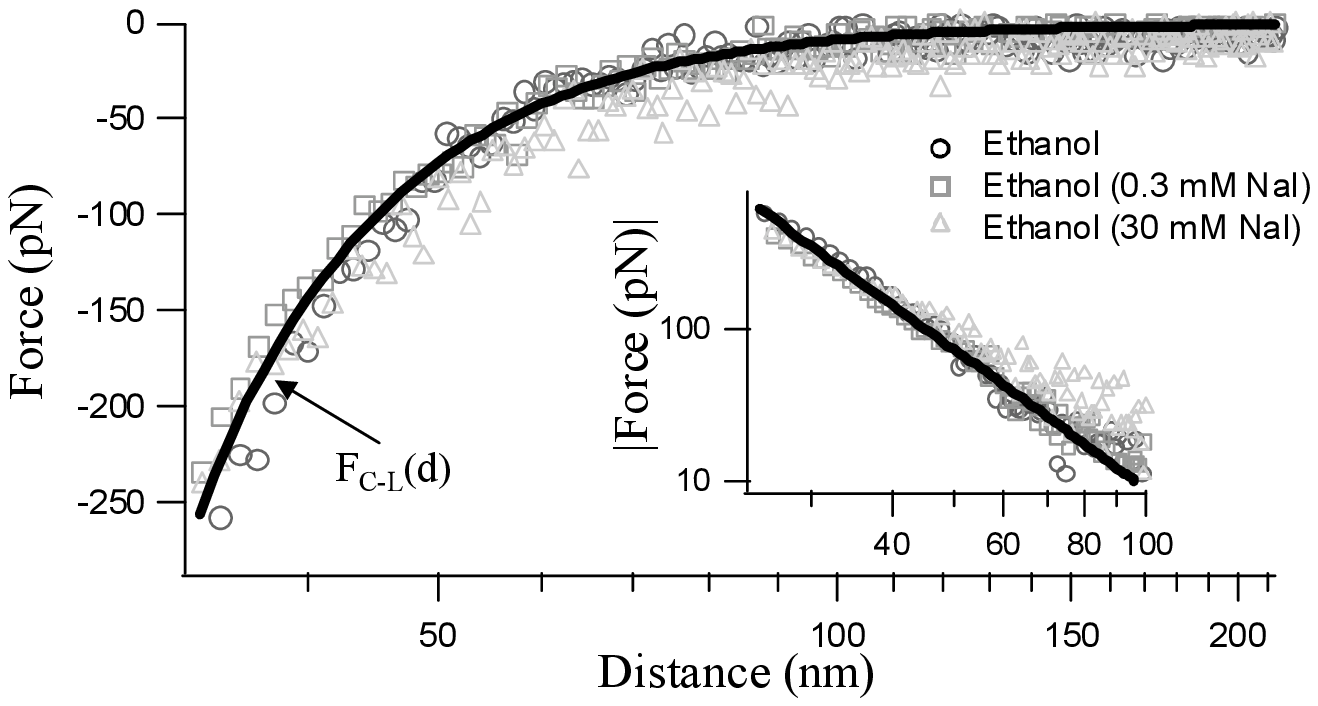}
		\caption{Comparison of the measured force with different concentrations of salt shows no significant difference. Force data in ethanol with no added salt (circles), 0.3 mM NaI (squares), and 30 mM NaI (triangles). Lifshitz's theory for no added salt is shown as a solid line. Inset: Log-log plot of the data.}
	\label{fig:ForceSalts}
\end{figure}	

The Casimir-Lifshitz force for different salt concentrations is shown in Fig.~\ref{fig:ForceSalts} along with Lifshitz's theory without corrections due to electrostatics or zero-frequency screening. The data is shown for experiments with no added salt (circles), 0.3 mM NaI (squares), and 30 mM NaI (triangles) and is obtained by averaging 51 data set for each concentration. The inset shows a log-log plot of the data. The difference between the forces due to the modification of the zero-frequency contribution and the Debye screening are greater for smaller separations and both are calculated to be $\sim$15 pN; however, the sensitivity of our apparatus is not adequate to distinguish a significant difference between these curves. While the averaged data follows the theoretical curve well, the standard deviation in the data is relatively large and varies from $\sim$90-130 pN in the range from 80-30 nm. Any variation in the three data sets (circles, squares, and triangles) is much less than the standard deviation and is not distinguishable in this experiment. In our previous work \cite{munday}, the standard deviation was $\sim$45-50 pN for separations of 80-50 nm, which is comparable to the error estimated in Ref. \cite{Davide} for Casimir force measurements in air obtained by considering the propagation of calibration errors. While the method presented in this paper reduces errors associated with drift and hydrodynamic forces over the previous method \cite{munday}, it has a larger standard deviation as a result of the subtraction of data from different force runs. Despite the large standard deviations, the averaged data is well described by Lifshitz's theory.
\section{CONCLUSIONS}
We have conducted detailed measurements of the Casimir-Lifshitz force in a fluid using an improved experimental setup and have considered various electrostatic contributions to the total force. NaI was added to ethanol to further reduce electrostatic interactions. No significant variation is found in the force upon modification of the grounding between the sphere and the plate or the increase of ionic concentration. Further improvements to the sensitivity are necessary to determine possible force differences at short range. Results are found to be consistent with Lifshitz's theory despite the relatively large uncertainties in both the experiment and the optical properties of the materials used.
\begin{acknowledgments}

The authors would like to acknowledge R. Podgornik, J. Zimmerberg, G. Carugno, and M. B. Romanowsky for helpful discussions. This project was partially supported by NSEC, 
under NSF Contract No. PHY-0117795, the Center for Nanoscale Systems at Harvard University, and the Intramural Research Program of the NIH, Eunice Kennedy Shriver National Institute of Child Health and Human Development. J.N.M. gratefully acknowledges financial support from the NSF. 

\end{acknowledgments}




\begin{thebibliography}{99}
\bibitem{AFMNote} Thermal drift between the sample and the tip is common in AFM measurements; however, for imaging of surfaces, the tip is near the surface and a feedback loop is used for compensation. Software is often used to remove this and additional artifacts.
\bibitem{WFNote} However, when the plastic holder is used, there are additional contributions to the electrostatic force that are not due to the work function differences.
\bibitem{DebyeNote} A Debye length of $\sim$24 nm was found for ethanol by assuming that the residue upon evaporation was due to a common salt, NaCl, see Ref.\cite{mundayR}. Here we obtain a Debye length of $\sim$100 nm through a linear extrapolation of the conductivity measurements for the ethanol used in these experiments.
\end{thebibliography}
\end{document}